\documentclass[aps,apl,twocolumn]{revtex4}

\usepackage{color}
\usepackage{graphicx}
\usepackage{natbib}
\usepackage{amsmath}
\usepackage{ulem}

\begin{document}

\newcommand{\bra}[1]{\langle #1|}
\newcommand{\ket}[1]{|#1\rangle}
\newcommand{\braket}[2]{\langle #1|#2\rangle}

\title{Direct Measurement of the Spatial-Spectral Structure of Waveguided Parametric Down-Conversion}
\author{Peter J. Mosley}
\affiliation{Max Planck Institute for the Science of Light, G\"unther-Scharowsky Strasse 1/Bau 24, 91058 Erlangen, Germany}
\author{Andreas Christ}
\affiliation{Max Planck Institute for the Science of Light, G\"unther-Scharowsky Strasse 1/Bau 24, 91058 Erlangen, Germany}
\email{Andreas.Christ@mpl.mpg.de}
\author{Andreas Eckstein}
\affiliation{Max Planck Institute for the Science of Light, G\"unther-Scharowsky Strasse 1/Bau 24, 91058 Erlangen, Germany}
\author{Christine Silberhorn}
\affiliation{Max Planck Institute for the Science of Light, G\"unther-Scharowsky Strasse  1/Bau 24, 91058 Erlangen, Germany}

\begin{abstract}
We present a study of the propagation of higher-order spatial modes in a waveguided parametric down-conversion photon pair source. Observing the multimode photon pair spectrum from a periodically poled KTiOPO$_4$ waveguide allowed us to isolate individual spatial modes through their distinctive spectral properties. We have measured directly the spatial distribution of each mode of the photon pairs, confirming the findings of our waveguide model, and demonstrated by coincidence measurements that the total parity of the modes is conserved in the nonlinear interaction. Furthermore, we show that we can combine the advantages of a waveguide source with the potential to generate spatially entangled photon pairs as in bulk crystal down-converters.
\end{abstract}

\maketitle

The prevalence of parametric down-conversion (PDC) as a source of photon pairs is due not only to the high quality of the photons produced but also its experimental simplicity relative to other methods of generating single photons. Its ubiquity may lead one to believe that PDC is a technique with little scope for improvement. However, bulk-crystal down-conversion sources suffer from a significant drawback: the photon pairs are emitted in a cone-shaped pattern making efficient collection difficult. This limits both the absolute count rate for a given pump power (stimulating the purchase of ever larger and more expensive laser systems) and, more importantly, the heralding efficiency of any bulk-crystal PDC source.

By confining photon pair generation to a channel  waveguide in a nonlinear optical material one can restrict the down-converted light to a well-defined set of spatial modes rather than allowing the emission to propagate at the natural phase-matching angles. This provides a straightforward method of controlling the messy spatial emission pattern from bulk-crystal down-converters and increases the down-conversion collection rate significantly \cite{URen2004Efficient-Conditional-Preparation, Chen2009A-versatile-waveguide-source, Tanzilli2001Highly-efficient-photon-pair, Fiorentino2007Spontaneous-parametric-down-conversion, Zhong2009High-performance-photon-pair}. However, as the wavelength of the down-converted pairs is approximately twice that of the pump, the light inevitably propagates not only in the fundamental waveguide mode but also in several higher-order spatial modes \cite{Roelofs1994Characterization-of-optical-waveguides, Banaszek2001Generation-of-correlated-photons, Karpinski2009Experimental-characterization-of-three-wave}. Because of the coupling between the spatial and spectral properties of the photon pairs imposed by phase-matching, these higher-order waveguide modes have a significant impact on the spectrum of the photon pairs and can markedly degrade source performance \cite{Eckstein2008Broadband-frequency-mode, Martin2009Integrated-optical-source}. Optimal source design requires that we both understand and control the interaction of higher-order modes in waveguides \cite{Christ2009Spatial-modes-in-waveguided}. On the other hand, the multimode spatial structure in the down-converted beams offers new opportunities for advanced quantum state preparation \cite{Saleh2009Modal-spectral-and-polarization}. Recent experiments utilize entanglement of the orbital angular momentum of the photon pairs prepared by bulk crystal PDC, but they rely on heavy filter operations by means of holographic state selection \cite{Mair2001Entanglement-of-the-orbital-angular, Arlt1999Parametric-down-conversion-for-light}. Waveguided PDC can directly provide entangled higher-order spatial modes in analogy to orbital angular momentum (OAM) entangled modes of photon pairs generated in bulk PDC experiments. OAM modes and their applications have been extensively studied recently \cite{Molina-Terriza2007Twisted-photons, Lassen2007Tools-for-Multimode-Quantum, Garcia-Escartin2008Quantum-multiplexing-with, Franke-Arnold2008Advances-in-optical-angular} with a view to to accessing higher-dimensional Hilbert spaces via the generation of hyperentangled photon pairs \cite{Barreiro2005Generation-of-Hyperentangled-Photon, Walborn2003Hyperentanglement-assisted-Bell-state-analysis, Franke-Arnold2002Two-photon-entanglement-of-orbital, Molina-Terriza2001Management-of-the-Angular-Momentum, Oemrawsingh2005Experimental-Demonstration-of-Fractional, Vaziri2002Experimental-Two-Photon-Three-Dimensional}. 

In this Letter we report the first direct observation of photon pairs generated in higher-order spatial modes by waveguided parametric down-conversion. We assign specific mode labels to each process by applying a numerical model and confirm parity conservation between the interacting mode triplets. Furthermore, we show that our source can generate spatially entangled two-photon states, while retaining the virtues of a waveguided device.

In general, down-converted photon pairs from waveguides are entangled in both frequency and spatial mode. Because of the dependence of the mode propagation constants on wavelength, spatial mode and spectrum are linked through the phase-matching conditions. Hence entanglement in these degrees of freedom cannot usually be separated \cite{Christ2009Spatial-modes-in-waveguided}. A key property of our source presented in this paper is its particular modal dispersion inside the waveguide which fulfills all the requirements to generate Bell-states in the spatial domain. By spectrally filtering the down-converted beams this source allows for the generation of photon pairs whose spatial entanglement is separated from the spectral domain. Hence hyperentangled photon pairs are emitted:
\begin{align}
    \left|\psi\right>_{\text{filtered}} =   B' \sum_k \lambda_k \ket{\psi^{(k)}_s, \phi^{(k)}_i}  \otimes \ket{\Psi}_{\text{Bell}}
    \label{eq:filteredPDCtheory}
\end{align}
with \(\ket{\psi^{(k)}_s}\) and \(\ket{\phi^{(k)}_i}\) denoting the spectral properties and \(\ket{\Psi}_{\text{Bell}}\) denoting a Bell state for higher-order spatial modes. Equation \,(\ref{eq:filteredPDCtheory}) is derived by applying a spectral Schmidt decomposition and introducing broadband frequency modes \(\ket{\psi^{(k)}_s}, \ket{\phi^{(k)}_i}\) to highlight the decoupling of the spectral and spatial degrees of freedom \cite{Rohde2007Spectral-structure-and-decompositions}.

Our source is a 10\,mm $z$-cut periodically-poled KTiOPO$_4$ (PPKTP) waveguide from AdvR with a nominal poling period of 8.72\,$\mu$m, pumped by a spatially filtered pulsed diode laser at 403.3\,nm with a bandwidth of 0.8 nm. Input coupling was through a 20$\times$ microscope objective and the pump was observed to be mainly (though not exclusively) in the fundamental mode of the waveguide. Output coupling was by an aspheric lens with a focal length of 6.24\,mm, set to image the output face of the waveguide to a plane about 800\,mm away. The type-II phase-matching conditions ensured that we obtained almost degenerate photon pairs, with the horizontally polarized pump ($y$-polarized in the crystal basis) yielding signal and idler with horizontal ($y$) and vertical ($z$) polarizations respectively. After the crystal the pump was removed with long-pass filters and the signal and idler photons were separated at a polarizing beamsplitter (PBS).

\begin{figure}
\includegraphics[width= 0.4 \textwidth]{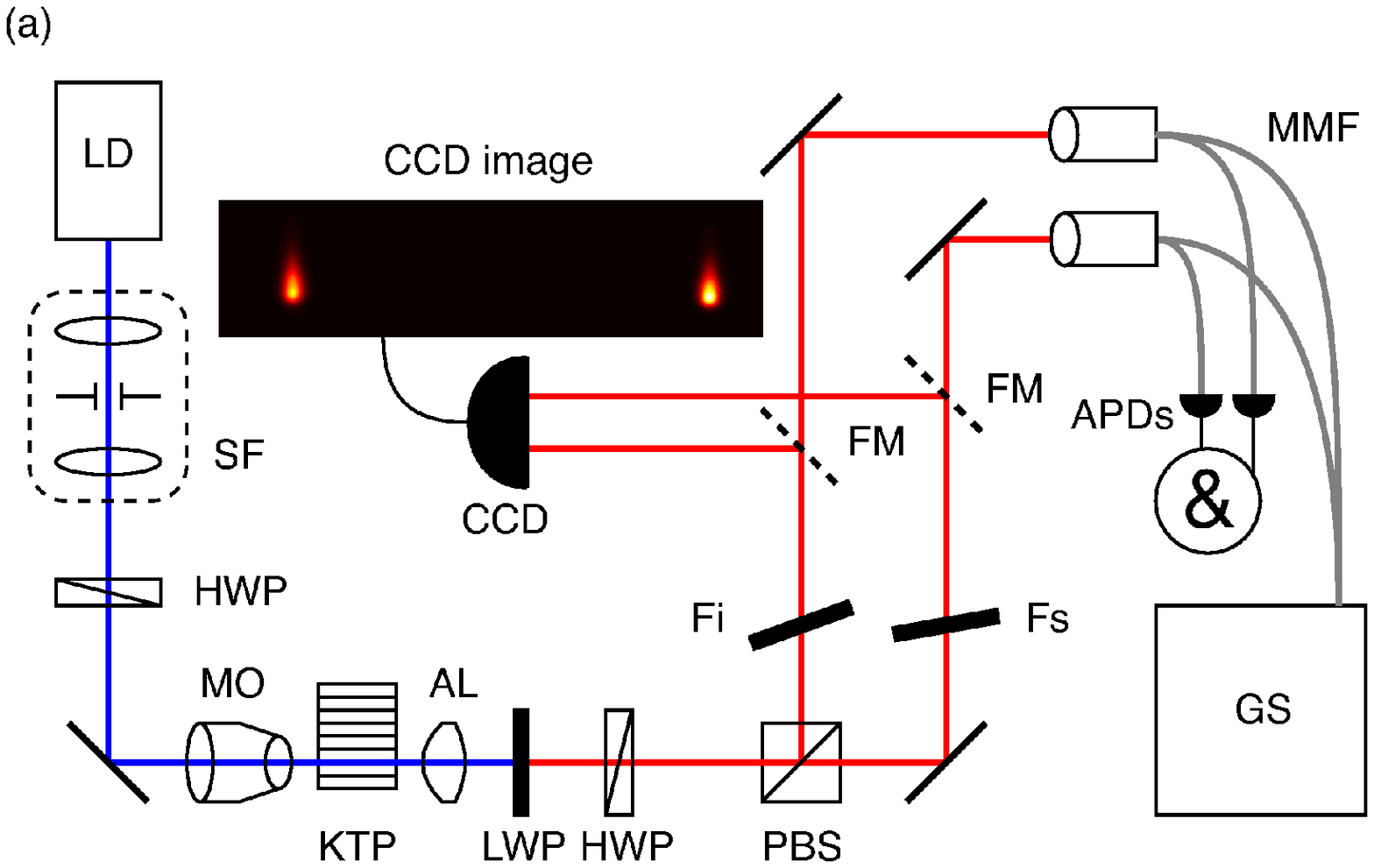}
\includegraphics[width= 0.48 \textwidth]{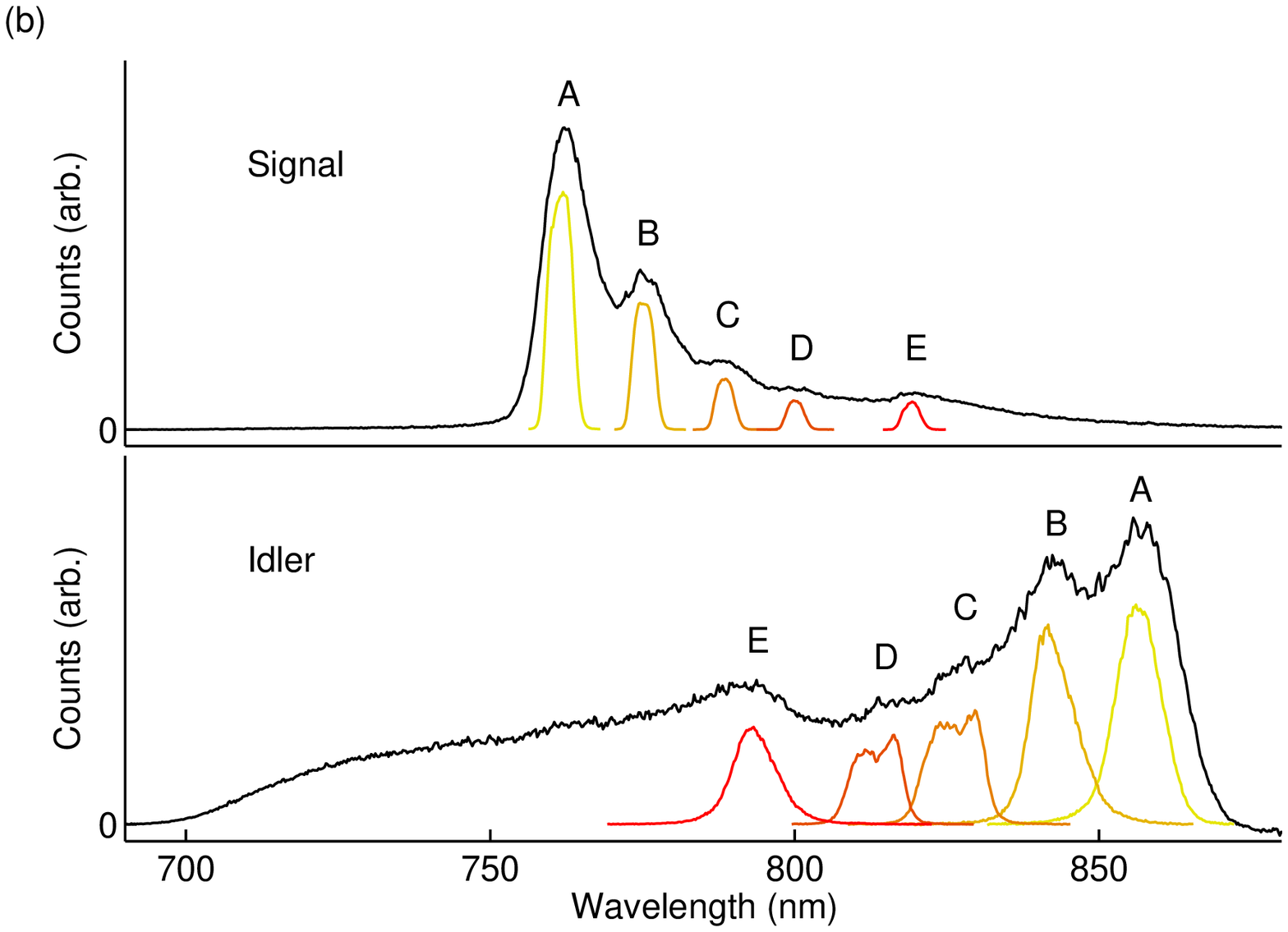}
\caption[Setup and spectra]{(a) Experimental setup. LD,laser diode; SF, spatial filter; HWP, half-wave plate; MO, microscope objective; KTP, PPKTP waveguide; AL, aspheric lens; LWP, long-wave pass filter; PBS, polarizing beamsplitter; Fs, Fi, interference filters; FM, flipper mirrors; MMF, multimode fiber; APD, avalanche photodiode; GS, grating spectrometer. (b) Background-subtracted spectra with correlated peaks labeled A to E. Colored lines show filtered spectral peaks (heights adjusted for ease of viewing).}
\label{fig:setup_spectra}
\end{figure}

Initially, signal and idler beams were coupled into two multimode fibers attached to a spectrometer with single-photon sensitivity. The multimode fibers allowed us to monitor simultaneously a range of spatial modes generated in the waveguide. The spectra for signal and idler are shown in Fig.\,\ref{fig:setup_spectra}. The spectral signatures of several spatial modes are clearly present: five individual peaks can be identified in both spectra. The peaks in the signal arm are paired with those in the idler through energy conservation. Each pair of peaks corresponds to a particular spatial mode set of pump, signal, and idler. Using the single-photon spectra, two sets of spectral filters (one for the signal with central wavelengths of 808\,nm and 830\,nm and bandwidths of 3\,nm and the other for the idler with central wavelengths 810, 830, and 860\,nm and bandwidths of 10\,nm) were calibrated such that each spectral peak could be individually selected by inserting and angle tuning a particular filter.

\begin{figure}
\includegraphics[width= 0.48 \textwidth]{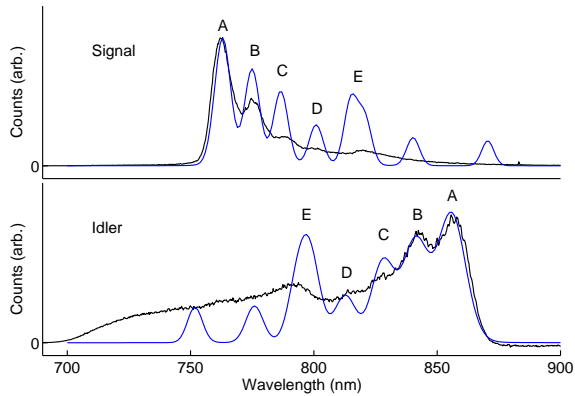}
\caption[Model spectra]{Comparison of modeled spectra with measured data. Labels A to E correspond to labels in Fig.\,\ref{fig:setup_spectra}.}
\label{fig:model_spectra}
\end{figure}

In order to assign specific mode labels to peaks A to E we developed a model of down-conversion in a step-index waveguide with a rectangular profile, bounded by a uniform dielectric on three sides and by air at the fourth. Although the production method of the waveguides results in a graded index distribution orthogonal to the air interface \cite{Fiorentino2007Spontaneous-parametric-down-conversion} our model yielded a simplified, semi-analytic solution which has been proven as sufficient to describe the experimental results \cite{Christ2009Spatial-modes-in-waveguided}. Although a more precise model would cause slight alterations to the predicted peak heights and spatial mode distributions, the more salient features --- the central wavelengths of the spectral peaks --- would remain virtually unchanged. By adjusting the index contrast, the waveguide dimensions, and the poling period as free parameters we fitted the calculated spectra to the measured marginal spectra of signal and idler (Fig.\,\ref{fig:model_spectra}). Note that the spectral widths of the signal and idler marginal distributions are set by the overlap of the pump bandwidth with the modal phase-matching functions of the crystal. Here the relatively broadband pump results in photon pairs with wider bandwidths than in similar experiments utilizing continuous-wave lasers \cite{Fiorentino2007Spontaneous-parametric-down-conversion}. From the theoretical modeling we identified each of the mode triplets listed in Table\,\ref{tab:modes}; they are labeled with the number of nodes in the horizontal and vertical directions $(x, y)$ respectively.

The principal mode pair A, is the result of the interaction between the fundamental modes of all three fields. This triplet has the most widely separated spectral components and was fitted by adjusting the poling period in the model. This effective poling period of $8.92\,\mu \textrm{m}$ serves as a global correction to allow for the difference between the waveguide in the lab and the empirical Sellmeier equations \cite{Kato2002Sellmeier-and-Thermo-Optic-Dispersion}. The remaining free parameters were adjusted to reproduce the observed marginal spectra of the photon pairs. With waveguide dimensions of $4.1 \times 9.3\,\mu \textrm{m}$ and an index contrast of 0.008 we obtained a very good agreement between theory and experiment (see Fig. \ref{fig:model_spectra}). These dimensions were verified under an optical microscope.

According to our model, mode pairs A, C, and E originate from the $(0,0)$ component of the pump (see Table \ref{tab:modes}). E stems from photon pairs generated both in modes $(1,0)$ and $(0,2)$ with overlapping spectral distributions. As a result of this frequency degeneracy and the coherence of the PDC process, the signal and idler pairs in E are entangled in spatial mode.  Further mode pairs occur at B and D pumped by the fraction of the pump intensity in the $(0,1)$ mode (37.5\%). In these cases, the signal and idler are in different --- though still parity conserving --- modes. Peaks B and D each consist of two down-conversion processes with almost identical spectra each entangled in spatial mode. The two pairs of peaks in the modeled spectra not mirrored in the measured data are from down-conversion events into higher order spatial modes [up to (1,2)]. These modes couple poorly into the collection fibers and hence are not seen in the data. The discrepancies in peak height between theory and experiment in Fig. \ref{fig:model_spectra} stem from our rectangular waveguide model and the falling collection efficiencies for higher-order spatial modes.

\begin{table}
\centering
\begin{tabular}{p{1cm} c c c c c c c c c} \hline \hline
A	&	$(0,0)_p$	&	$\rightarrow$	&	$(0,0)_s$	&	+	&	$(0,0)_i$	&	&	&	&	\\ 
B	&	$(0,1)_p$	&	$\rightarrow$	&	$(0,0)_s$	&	+	&	$(0,1)_i$	&	and	&	$(0,1)_s$	&	+	&	$(0,0)_i$	\\ 
C	&	$(0,0)_p$	&	$\rightarrow$	&	$(0,1)_s$	&	+	&	$(0,1)_i$	\\ 
D	&	$(0,1)_p$	&	$\rightarrow$	&	$(0,1)_s$	&	+	&	$(0,2)_i$	&	and	&	$(0,2)_s$	&	+	&	$(0,1)_i$	\\ 
E	&	$(0,0)_p$	&	$\rightarrow$	&	$(1,0)_s$	&	+	&	$(1,0)_i$	&	and	&	$(0,2)_s$	&	+	&	$(0,2)_i$	\\ \hline \hline
\end{tabular}
\caption[Mode designations]{Processes giving rise to the five observed mode spectra.}
\label{tab:modes}
\end{table}

\begin{figure*}
\includegraphics[height= 0.23 \textheight]{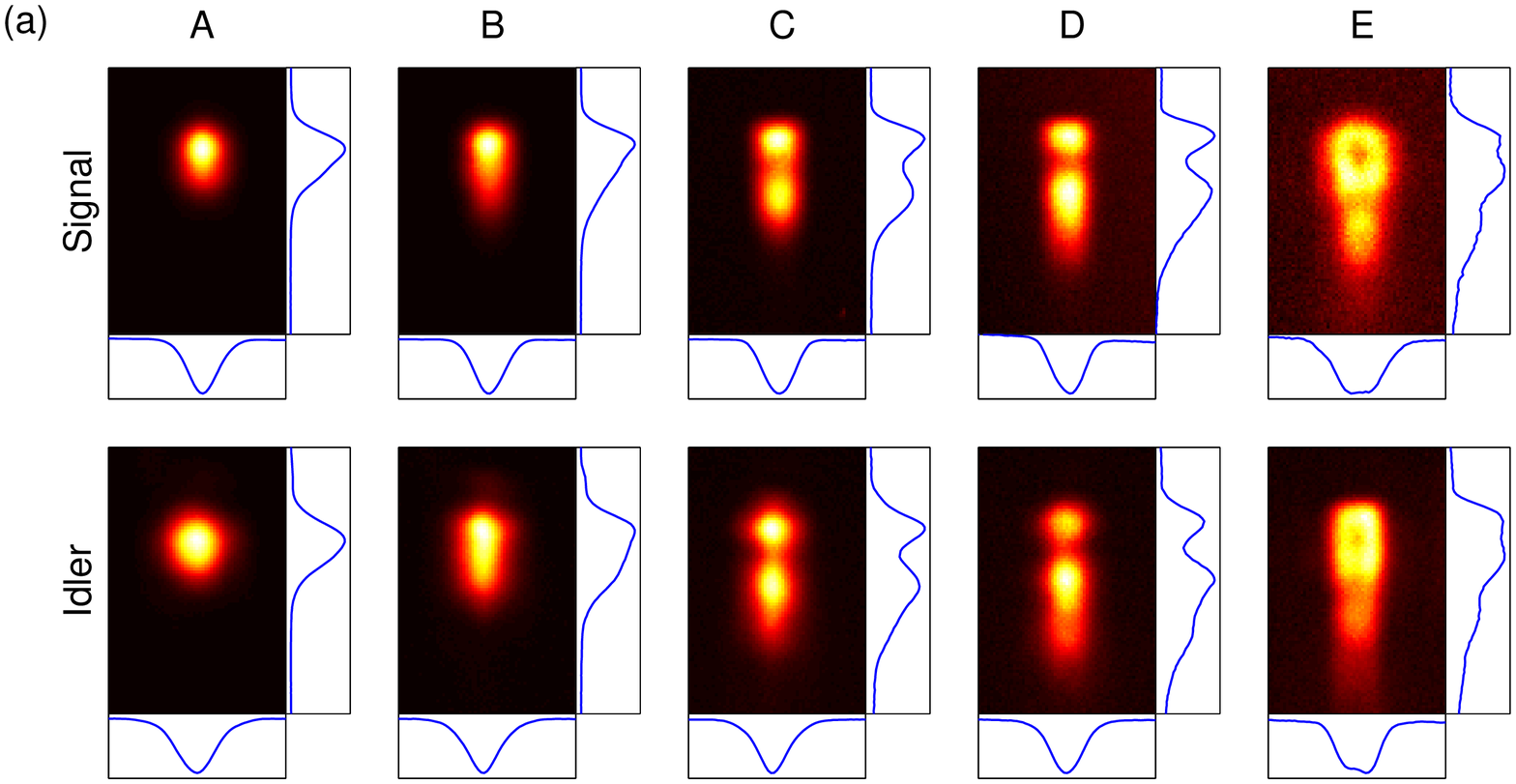}\includegraphics[height= 0.23 \textheight]{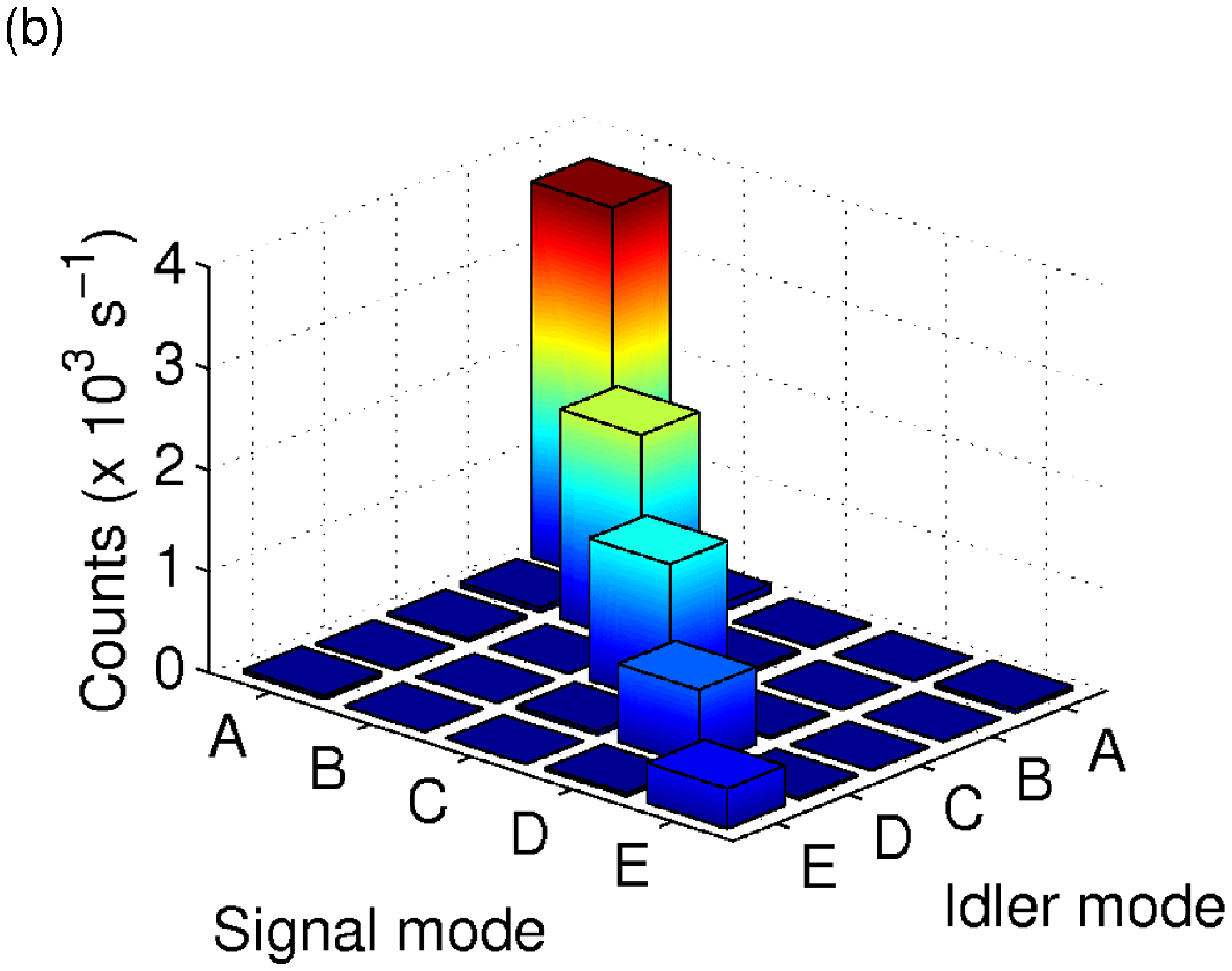}
\caption[Spatial modes]{(a) Plots of the background-subtracted spatial mode distributions for signal (top) and idler (bottom) arms along with their marginal distributions. The air interface is towards the top of each frame, and the decay of the waveguide structure into the chip can clearly be observed towards the bottom. (b) Coincidence count rate between different pairs of spatial modes. Labels A to E correspond to labels in Fig.\,\ref{fig:setup_spectra}.}
\label{fig:mode_coinc}
\end{figure*}

Next the high-sensitivity CCD camera was removed from the spectrometer and placed in the image plane of the $f$\,=\,6.24\,mm aspheric to measure the spatial intensity distribution of each mode. Both output beams from the waveguide were directed simultaneously to separate areas of the sensor yielding magnified images (approximately 130$\times$) of the spatial modes of both signal and idler in the waveguide. Fig.\,\ref{fig:mode_coinc} shows the characteristic distributions of individual spatial modes, recorded by tuning the spectral filters to pick out spatial modes through their unique spectra. This demonstrates the strong correlation between the spatial and spectral degrees of freedom in this system.

In these measurements, a high level of background was present from the long-lived, unphasematched fluorescence emitted by the waveguide that could not be removed by time-gating due to the slow speed of the camera. Instead the fluorescence level was measured for each spatial mode by rotating the pump polarization to vertical, hence removing any phase-matching. However, this background could not be subtracted directly as the fluorescence was higher for a vertically polarized pump. Therefore, auxiliary measurements for both polarizations were made with an unpoled waveguide in which no phasematched processes could take place. The ratio between these fluorescence signals allowed us to introduce a correction for the scaling of the background in the PDC spatial mode images recorded with the poled waveguide. Subtracting this adapted background from the spatial mode images yielded a realistic measurement of the true distribution of the PDC in the various spatial modes as shown in Fig.\,\ref{fig:mode_coinc}.

It is apparent from Fig.\,\ref{fig:mode_coinc} that each of the five mode pairs A to E has its own characteristic spatial intensity distribution, with peaks B to E exhibiting obvious signs of higher-order mode propagation. All of the recorded spatial distributions agree very well with those found in the spectral degree of freedom through the model as listed in Table \ref{tab:modes}: A and C are pure $(0,0)$ and $(0,1)$ respectively; B is a sum of $(0,0)$ and $(0,1)$ where two processes overlap spectrally; D is also a sum of two processes, $(0,1)$ and $(0,2)$; E, the superposition of the $(1,0)$ and $(0,2)$ modes, is the only case to show a higher-order mode in the horizontal direction. Furthermore one can see the deviation of the waveguide from a rectangular structure: the fundamental mode sits at the top of the guide close to the air boundary, while higher-order modes spread down into the chip where there is an exponential decay in the refractive index contrast not present in our model.

Finally, we performed a coincidence measurement between the different spatial modes. With both beams once again coupled into the multimode fibers the photons were sent to two silicon avalanche photodiodes (APD). The time-gated single count rate of each APD was monitored along with the rate of coincidence counts between the two as the filters were set to select every combination of the five spatial modes in both the signal and idler arms. The results are shown in Fig.\,\ref{fig:mode_coinc} with the background of accidental coincidences --- calculated as the product of the singles rates divided by the laser repetition rate (1\,MHz) --- subtracted from the coincidence rates. The presence of only diagonal elements in the corrected coincidence rates demonstrates the strict correlation between the spatial modes: if the signal photon is emitted into a particular spatial mode then the idler will always be found in the corresponding mode. This confirms the requirement for parity conservation between the three interacting modes.

Our measurements demonstrate that the generation of higher-order spatial mode entanglement can be easily accomplished in waveguided PDC. For our source this can be achieved by filtering processes B, D, or E and postselecting on successful coincidence events. For example, by choosing only process B we find
\vspace{-0.18cm}
\begin{multline}
    \ket{\psi}_{\text{B}} =   B' \sum_k \lambda_k \ket{\psi^{(k)}_s, \phi^{(k)}_i} \\
    \otimes \underbrace{\left( \ket{ (0,1)_s, (0,0)_i} + \ket{ (0,0)_s, (0,1)_i} \right)}_{\ket{\Psi^+}}.
    \label{eq:filteredPDC}
\end{multline}
Similarly, filtering peaks D and E yields the Bell states \(\ket{\Psi^+}\) and \(\ket{\Phi^+}\) respectively.

In conclusion, we have directly imaged spectrally-resolved spatial modes of PDC in a PPKTP waveguide. We have identified the individual spatial mode contributions and demonstrated that our model accurately reproduces the photon pair spectra. This shows that waveguided PDC sources have the potential to be used as bright sources of photon pairs entangled in spatial mode. These photon pairs may have many applications from testing the Bell inequalities in the spatial domain to distributing Bell pairs over free space links for quantum key distribution applications.

This work was supported by the EC under the FET-Open grant agreement CORNER, number FP7-ICT-213681.

\end{document}